\documentstyle[12pt,epsfig]{article}
\begin{document}
\begin{center}
\begin{Large}
{\bf  Search for New Particles Decaying to $b\bar{b}$ in $p\bar{p}$ Collisions 
at $\sqrt{s}=1.8$ TeV}
\end{Large}
\end{center}
\font\eightit=cmti8
\def\r#1{\ignorespaces $^{#1}$}
\hfilneg
\begin{sloppypar}
\noindent
F.~Abe,\r {17} H.~Akimoto,\r {39}
A.~Akopian,\r {31} M.~G.~Albrow,\r 7 A.~Amadon,\r 5 S.~R.~Amendolia,\r {27} 
D.~Amidei,\r {20} J.~Antos,\r {33} S.~Aota,\r {37}
G.~Apollinari,\r {31}
% ...
\end{sloppypar}
\vskip .026in
\begin{center}
(CDF Collaboration)
\end{center}

\vskip .026in
\begin{center}
\r 1  {\eightit Argonne National Laboratory, Argonne, Illinois 60439} \\
\r 2  {\eightit Istituto Nazionale di Fisica Nucleare, University of Bologna,
I-40127 Bologna, Italy} \\
\r 3  {\eightit Brandeis University, Waltham, Massachusetts 02254} \\
\r 4  {\eightit University of California at Los Angeles, Los 
Angeles, California  90024} \\  
\r 5  {\eightit University of Chicago, Chicago, Illinois 60637} \\
\r 6  {\eightit Duke University, Durham, North Carolina  27708} \\
\r 7  {\eightit Fermi National Accelerator Laboratory, Batavia, Illinois 
60510} \\
\r 8  {\eightit University of Florida, Gainesville, Florida  32611} \\
\r 9  {\eightit Laboratori Nazionali di Frascati, Istituto Nazionale di Fisica
               Nucleare, I-00044 Frascati, Italy} \\
\r {10} {\eightit University of Geneva, CH-1211 Geneva 4, Switzerland} \\
\r {11} {\eightit Harvard University, Cambridge, Massachusetts 02138} \\
\r {12} {\eightit Hiroshima University, Higashi-Hiroshima 724, Japan} \\
\r {13} {\eightit University of Illinois, Urbana, Illinois 61801} \\
\r {14} {\eightit Institute of Particle Physics, McGill University, Montreal 
H3A 2T8, and University of Toronto,\\ Toronto M5S 1A7, Canada} \\
\r {15} {\eightit The Johns Hopkins University, Baltimore, Maryland 21218} \\
\r {16} {\eightit Institut f\"{u}r Experimentelle Kernphysik, 
Universit\"{a}t Karlsruhe, 76128 Karlsruhe, Germany} \\
\r {17} {\eightit National Laboratory for High Energy Physics (KEK), Tsukuba, 
Ibaraki 305, Japan} \\
\r {18} {\eightit Ernest Orlando Lawrence Berkeley National Laboratory, 
Berkeley, California 94720} \\
\r {19} {\eightit Massachusetts Institute of Technology, Cambridge,
Massachusetts  02139} \\   
\r {20} {\eightit University of Michigan, Ann Arbor, Michigan 48109} \\
\r {21} {\eightit Michigan State University, East Lansing, Michigan  48824} \\
\r {22} {\eightit University of New Mexico, Albuquerque, New Mexico 87131} \\
\r {23} {\eightit The Ohio State University, Columbus, Ohio  43210} \\
\r {24} {\eightit Osaka City University, Osaka 588, Japan} \\
\r {25} {\eightit Universita di Padova, Istituto Nazionale di Fisica 
          Nucleare, Sezione di Padova, I-35131 Padova, Italy} \\
\r {26} {\eightit University of Pennsylvania, Philadelphia, 
        Pennsylvania 19104} \\   
\r {27} {\eightit Istituto Nazionale di Fisica Nucleare, University and Scuola
               Normale Superiore of Pisa, I-56100 Pisa, Italy} \\
\r {28} {\eightit University of Pittsburgh, Pittsburgh, Pennsylvania 15260} \\
\r {29} {\eightit Purdue University, West Lafayette, Indiana 47907} \\
\r {30} {\eightit University of Rochester, Rochester, New York 14627} \\
\r {31} {\eightit Rockefeller University, New York, New York 10021} \\
\r {32} {\eightit Rutgers University, Piscataway, New Jersey 08855} \\
\r {33} {\eightit Academia Sinica, Taipei, Taiwan 11530, Republic of China} \\
\r {34} {\eightit Texas A\&M University, College Station, Texas 77843} \\
\r {35} {\eightit Texas Tech University, Lubbock, Texas 79409} \\
\r {36} {\eightit Istituto Nazionale di Fisica Nucleare, University of Trieste/
Udine, Italy} \\
\r {37} {\eightit University of Tsukuba, Tsukuba, Ibaraki 315, Japan} \\
\r {38} {\eightit Tufts University, Medford, Massachusetts 02155} \\
\r {39} {\eightit Waseda University, Tokyo 169, Japan} \\
\r {40} {\eightit University of Wisconsin, Madison, Wisconsin 53706} \\
\r {41} {\eightit Yale University, New Haven, Connecticut 06520} \\
\end{center}

\renewcommand{\baselinestretch}{2}
\large
\normalsize

\begin{center}
{\bf Abstract}
\end{center}
We have used 87 ${\rm pb}^{-1}$ of data collected with the Collider Detector 
at Fermilab to search for new particles decaying to $b\bar{b}$. We present 
model--independent upper limits on the cross section for narrow resonances
which exclude the color--octet technirho in the mass interval
$350 < M < 440$ GeV/c$^2$.  
In addition, we exclude topgluons, 
predicted in models of topcolor--assisted technicolor, of width 
$\Gamma=0.3 M$ in the mass range $280 < M < 670 $ GeV/c$^2$, 
of width $\Gamma = 0.5 M$ in the mass range $340 < M < 640$ GeV/c$^2$, and  
of width $\Gamma = 0.7 M$ in the mass range $375 < M < 560$ GeV/c$^2$.

\vspace*{0.5in}

\clearpage

In this paper we report on a search for new particles decaying to $b\bar{b}$.
In addition to a model independent search for narrow resonances, we perform a 
specific search for resonances from topcolor--assisted 
technicolor~\cite{ref_topcolor} . 
Since electroweak symmetry breaking is associated with the origin of fermion
masses, the large mass of the top quark suggests that the third generation
could contain clues about the origin of electroweak symmetry 
breaking. Topcolor--assisted technicolor is a model in which the top quark is 
heavy because of a new interaction. Topcolor replaces the $SU(3)_C$ of QCD with 
$SU(3)_1$ which couples only to 
the first two quark generations and $SU(3)_2$ which couples only to the third 
generation. The symmetry $SU(3)_1 \times SU(3)_2$ is broken, resulting in 
the familiar $SU(3)_C$ of QCD and an additional SU(3) which couples 
mainly to the third generation.
In addition, there is a U(1) symmetry added to keep the $b$ quark light.
The additional SU(3) symmetry gives rise to a color--octet gauge
boson, the topgluon $g_T$, while the additional U(1) symmetry
gives rise to  a new heavy neutral gauge boson, the topcolor
$Z'_T$. Topgluons are expected to have a large width, so we search for three different
widths $\Gamma=0.3M$, $0.5M$, and $0.7M$, where $M$ is the new particle's mass.
For narrow resonances we consider color--octet technirhos from a model of 
walking technicolor~\cite{ref_lane}, a $Z^{\prime}$ from topcolor--assisted
technicolor~\cite{ref_topcolor}, a $Z^{\prime}$ with standard model 
couplings~\cite{ref_zprime}, and a vector bound state of 
gluinos (vector gluinonium) appearing in supersymmetry models ~\cite{ref_susy}. 
We search for these new phenomena in the $b\bar{b}$ mass spectrum in $p\bar{p}$
collisions at a center of mass energy $\sqrt{s}=1.8$ TeV.

A detailed description of the Collider Detector at Fermilab (CDF) can be found 
elsewhere~\cite{ref_CDF}. We use a coordinate system with the $z$ axis along the proton
beam, transverse coordinate perpendicular to the beam, azimuthal angle $\phi$, 
polar angle $\theta$, and pseudorapidity $\eta=-\ln \tan(\theta/2)$. 
The silicon vertex detector (SVX), a four-layer silicon strip device with 
radiation--hard electronics, located immediately outside the beampipe, provides 
precise track reconstruction in the transverse plane and is used to 
identify secondary--vertices from $b$ quark decays.  The momenta of charged 
particles are measured in the Central Tracking Chamber (CTC), which is inside
a 1.4 T superconducting solenoidal magnet. Outside the CTC, electromagnetic
and hadronic calorimeters, segmented in $\eta$-$\phi$ towers, cover the 
pseudorapidity region $|\eta|<4.2$.

Jets are reconstructed as localized energy depositions in the CDF calorimeters.
The jet energy $E$ is defined as the scalar sum of the calorimeter tower 
energies inside a cone of radius 
$R=\sqrt{(\Delta\eta)^2 + (\Delta\phi)^2}=0.7$, centered on the jet direction. 
The jet momentum $\vec{P}$ is the corresponding vector sum: 
$\vec{P} = \sum{E_i\hat{u}_i}$ with $\hat{u}_i$ being the 
unit vector pointing from the interaction point to the energy deposition $E_i$ 
in tower $i$ inside the same cone. $E$ and $\vec{P}$ are corrected 
for calorimeter non-linearities, energy lost in uninstrumented regions of the
detector and outside the clustering cone, and
energy gained from the underlying event and multiple $p\bar{p}$ interactions. 
Full details of jet reconstruction and
jet energy corrections at CDF can be found elsewhere~\cite{ref_jet}.

We define the dijet system as the
two jets with the highest transverse momentum in an event (leading jets) 
and define the dijet mass
$m=\sqrt{(E_1 + E_2)^2 - (\vec{P}_1 + \vec{P}_2)^2}$.  The dijet mass
resolution is approximately 10\% for dijet masses above 150 GeV/c$^2$.
Our data sample was obtained using four triggers
that required at least one jet with uncorrected cluster transverse energies
of 20, 50, 70 and 100 GeV, respectively.  After jet energy corrections these 
trigger samples
were used to measure the dijet mass spectrum above 150, 217, 292 and 388 
GeV/c$^2$, respectively.  At these mass thresholds the corresponding trigger 
efficiencies were greater than 93\%. The four data samples 
corresponded to integrated luminosities of $0.087$, $2.2$, $11$ and $87$ 
pb$^{-1}$ after prescaling.  
We selected events with two or more jets and 
required that the two leading jets have pseudorapidity $|\eta_1|<2$ and
$|\eta_2|<2$ and a scattering angle in the dijet center-of-mass frame 
$|\cos\theta^*| = |\tanh[(\eta_1-\eta_2)/2]| < 2/3$. 
The $\cos\theta^*$ 
requirement ensures uniform acceptance as a function of mass and reduces the 
QCD background which  peaks at $|\cos\theta^*|=1$.  
To maintain the projective nature of the calorimeter towers, the $z$ position 
of the event vertex was required to be within 60 cm of the 
center of the detector; this cut removed 7\% of the events. 
Backgrounds from cosmic rays, 
beam halo, and detector noise were removed 
by requiring \mbox{${\not\!\!E_T}/\sqrt{\sum E_T}<6$} GeV$^{1/2}$ and 
$\sum E< 2$ TeV,
where \mbox{${\not\!\!E_T}$} is the missing transverse
energy, $\sum E_T$ is the total transverse energy, and $\sum E$ is the total
energy in the event.

To identify jets originating with a $b$--quark we require that tracks reconstructed
in the CTC and SVX form a secondary--vertex, displaced from the event vertex.
In the secondary--vertex algorithm, described elsewhere~\cite{ref_top_prl}, we 
tightened the transverse momentum ($p_T$) requirements on the tracks as a 
function of dijet mass to reduce backgrounds at high mass~\cite{ref_kara}. 
The highest $p_T$ track in a reconstructed vertex was
required to have a $p_T$ of at least 2 GeV/c for dijets with a mass less 
than 321 GeV/c$^2$. This cut was tightened incrementally as a function of 
dijet mass to 5 GeV/c for dijets with a mass greater than 470 GeV/c$^2$.
For secondary--vertices formed from only two tracks, we additionally required 
that the tracks have a minimum $p_T$ of at least 1 GeV/c for dijets with mass 
less than 321 GeV/c$^2$, rising to 2 GeV/c for dijets with mass greater than 
388 GeV/c$^2$. We required both jets
in the dijet to have a displaced secondary--vertex (a $b$--tag). The efficiency of 
our $b$--tagging algorithm was determined from a Monte Carlo simulation of 
detector response to the decay of a heavy object to $b\bar{b}$. The simulation 
was tuned to reproduce the observed tracking efficiency~\cite{ref_kara}.
The efficiency for $b$--tagging a heavy object decaying to $b\bar{b}$ 
decreases from 11\% to 2.5\% as the dijet mass increases from 200 to 650 
GeV/c$^2$. The efficiency decreases as the dijet mass increases because the 
secondary--vertex resolution and acceptance degrade as the distance from the 
event vertex increases and because the tracking efficiency degrades as the 
density of tracks within a jet increases.

\begin{figure}[tbh]
\centerline{\epsfig{figure=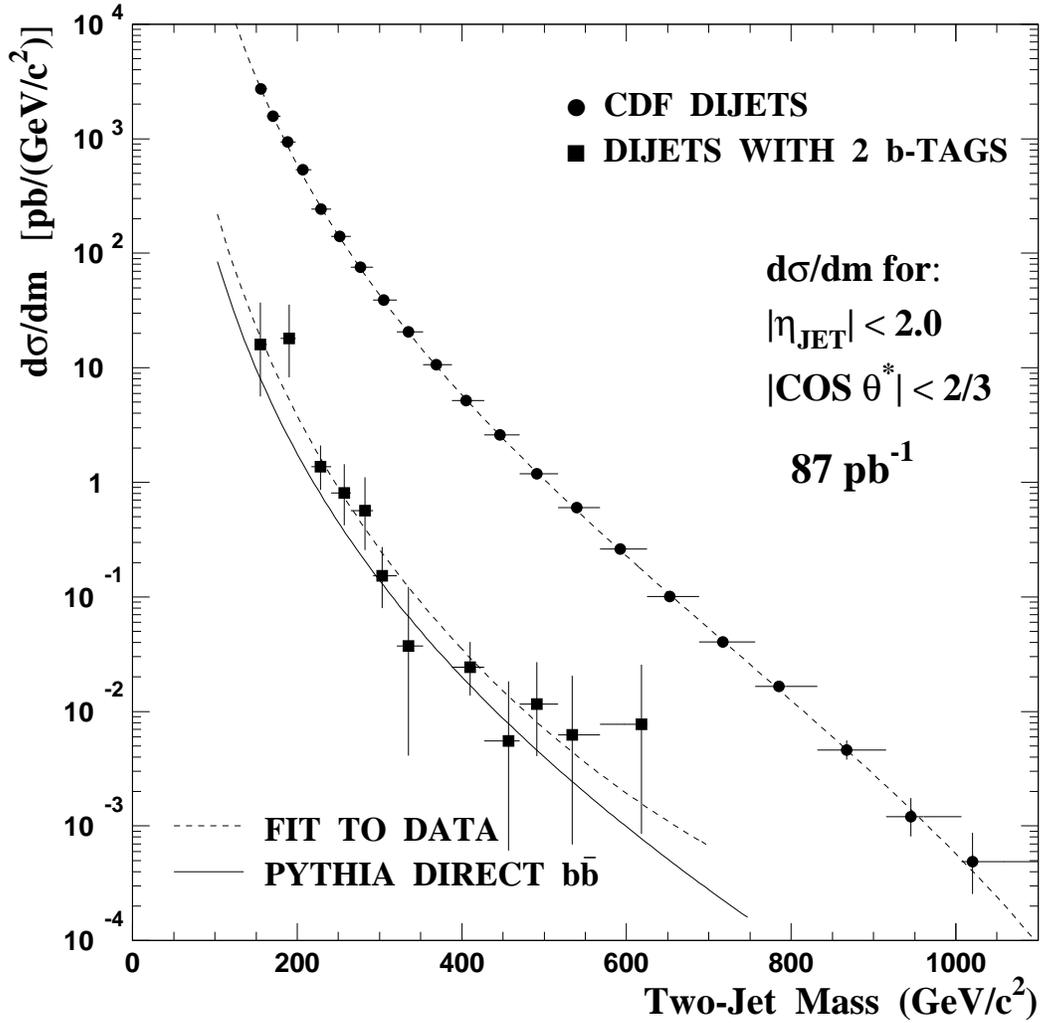,width=6in}}
\caption{ 
The dijet mass distribution (circles) and double--$b$--tagged dijet mass spectrum
(boxes) compared to a fit to a smooth parameterization (dashed curves). 
Also shown is a QCD prediction for $b\bar{b}$ production (solid curve).}
\label{fig_log}
\end{figure}

In Fig.~\ref{fig_log} we present the inclusive dijet mass 
distribution for untagged and double--$b$--tagged dijets with $|\eta|<2$ and 
$|\cos\theta^*|<2/3$.
The mass distributions have been corrected for the trigger, $z$-vertex, and 
$b$--tagging inefficiencies previously discussed.
We plot the differential cross section versus the mean dijet mass, $m$, in 
bins of width approximately equal to the mass resolution (RMS$\sim 10$\%). 
The $b$--tagged data are compared to a smooth parameterization and to a 
QCD prediction. The QCD prediction is for direct $b\bar{b}$ 
production from the PYTHIA Monte Carlo~\cite{ref_pythia}, with the $b$ quarks
decayed by the CLEO Monte Carlo QQ, and includes a simulation of the 
CDF detector. 
Direct $b\bar{b}$ production includes the processes 
$q\bar{q}\rightarrow b\bar{b}$ and $gg\rightarrow b\bar{b}$; other $b\bar{b}$ 
production processes do not contribute significantly.
The QCD simulation used CTEQ2L parton distributions~\cite{ref_CTEQ} and a 
renormalization scale $\mu=P_T$.

\begin{figure}[tbh]
\centerline{\epsfig{figure=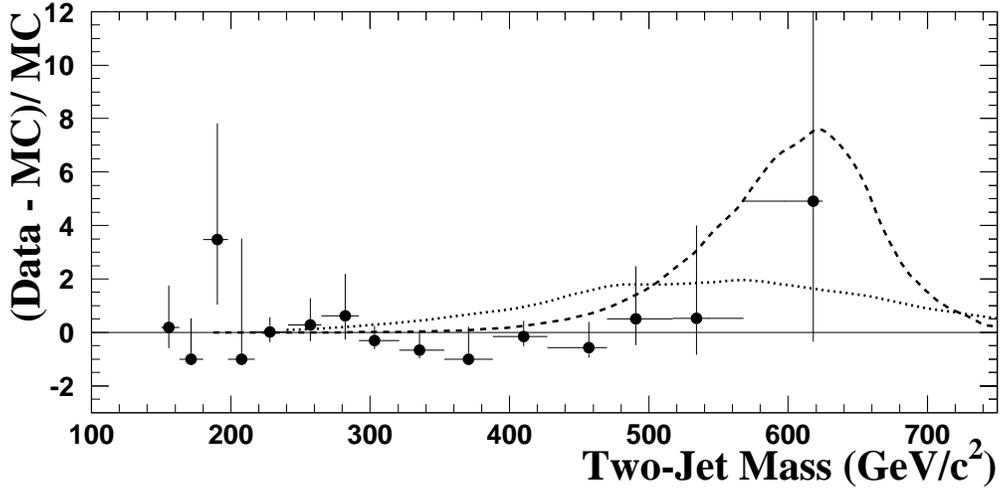,width=6in}}
\vspace*{-2.5in}
\caption{ 
The fractional difference between the double--$b$--tagged dijet mass distribution 
(points) and
a background prediction (solid line) is compared to a simulation of 
a 600 GeV/c$^2$ narrow resonance (dashed curve) and a 600 GeV/c$^2$ topgluon of width
$\Gamma=0.5$M (dotted curve) in the CDF detector.
The background has been normalized to fit the data and the resonances have 
each been normalized to the 95\% CL upper limit on the cross section 
for a 600 GeV/c$^2$ resonance.}
\label{fig_lin}
\end{figure}

To search for resonances, we fit the data to the shape of the $b\bar{b}$
Monte Carlo calculation and a new particle resonance.  The 
$b\bar{b}$ prediction multiplied by a factor of $1.7$ fits the data well 
($\chi^2/$DF$=0.61$), as shown in 
Fig.~\ref{fig_lin} which also shows the predicted line shape for narrow 
resonances and topgluons. Narrow resonances were modeled with a PYTHIA 
simulation of a $Z^{\prime}$ decaying to $b\bar{b}$ followed by the CDF 
detector simulation. The mass resolution is 
dominated by a Gaussian distribution from jet energy resolution and a long 
tail towards low mass from QCD radiation. 
Since the natural width of the $Z^{\prime}$ is significantly smaller than the 
reconstructed width, these mass resonance curves were used to model the shape of 
all narrow resonances decaying to $b\bar{b}$.
Topgluons were modeled using a PYTHIA simulation, in which we inserted the
parton-level sub-process cross section for topgluons~\cite{ref_lane}, followed
by the CDF detector simulation.  We simulated topgluons with width
$\Gamma = 0.3$M, $0.5$M, and $0.7$M.
There is no evidence in the data for either narrow or wide resonances decaying 
to $b\bar{b}$.

\begin{figure}[tbh]
\centerline{\epsfig{figure=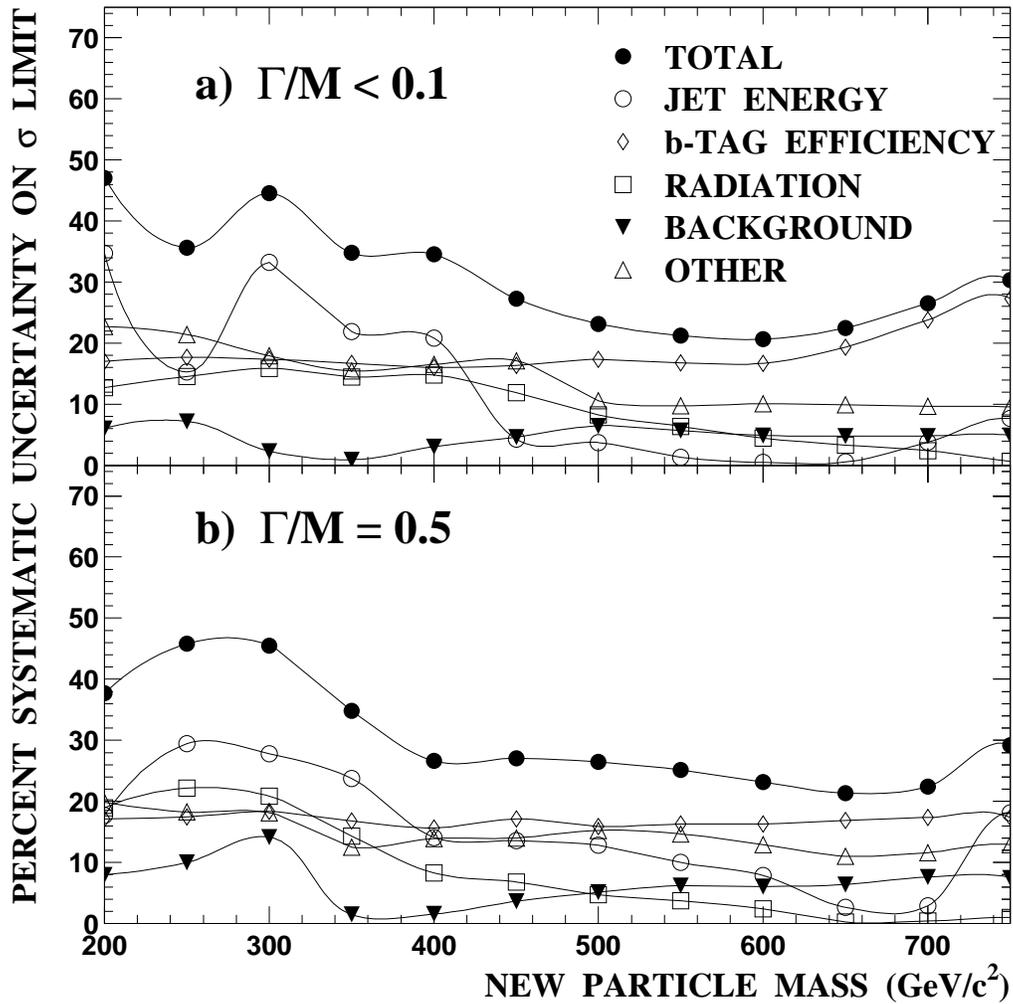,width=6in}}
\caption{ 
Systematic uncertainties on the cross section upper limit for a) narrow 
resonances and b) topgluons of width $\Gamma=0.5$M as a function of mass.}
\label{fig_sys}
\end{figure}

Systematic uncertainties on the cross section for observing a new particle 
in the CDF detector are shown in Fig.~\ref{fig_sys}.  
Each systematic uncertainty on the fitted signal cross section was determined 
by varying the source of uncertainty by $\pm 1\sigma$ and refitting.
The sources of uncertainty presented in Fig.~\ref{fig_sys} are the jet energy 
scale uncertainty, the $b$--tagging efficiency, 
the effect of QCD radiation on the mass resonance line 
shape, the shape of the $b\bar{b}$ background predicted by QCD, and other 
sources including trigger efficiency, jet energy resolution for narrow 
resonances, relative jet 
energy corrections between different parts of the CDF calorimeter, and 
luminosity. The dominant systematic uncertainty 
at low mass is a 3\% uncertainty in the jet energy scale.
The dominant systematic uncertainty at high mass was the 
uncertainty in the $b$--tagging efficiency, which varied from 14\% at low mass 
to 24\% at high mass.  The uncertainty on the shape of the $b\bar{b}$ 
background, from using various parton distributions, produced a relatively 
small uncertainty in the cross section limits compared to the dominant sources
discussed above, which primarily affect the signal.
Other possible sources of background, such as mistags or charm, may contribute
to the normalization but are expected to have a similar shape to $b\bar{b}$, 
and do not contribute significantly to the systematic uncertainty.
The total systematic uncertainty was found by adding the individual 
sources in quadrature. 

In the absence of evidence for new physics we proceeded to set upper
limits on the cross section for new particles. 
For each value of new particle mass in 50 GeV/c$^2$ steps from 200 to 750
GeV/c$^2$, we performed a binned maximum likelihood fit of the data to the 
background shape and the mass resonance shape. 
We convoluted each of the Poisson likelihood distributions with the 
corresponding 
total Gaussian systematic uncertainty, and found the 95\% confidence level 
(CL) upper limit presented in Table I.

\begin{figure}[tbh]
\centerline{\epsfig{figure=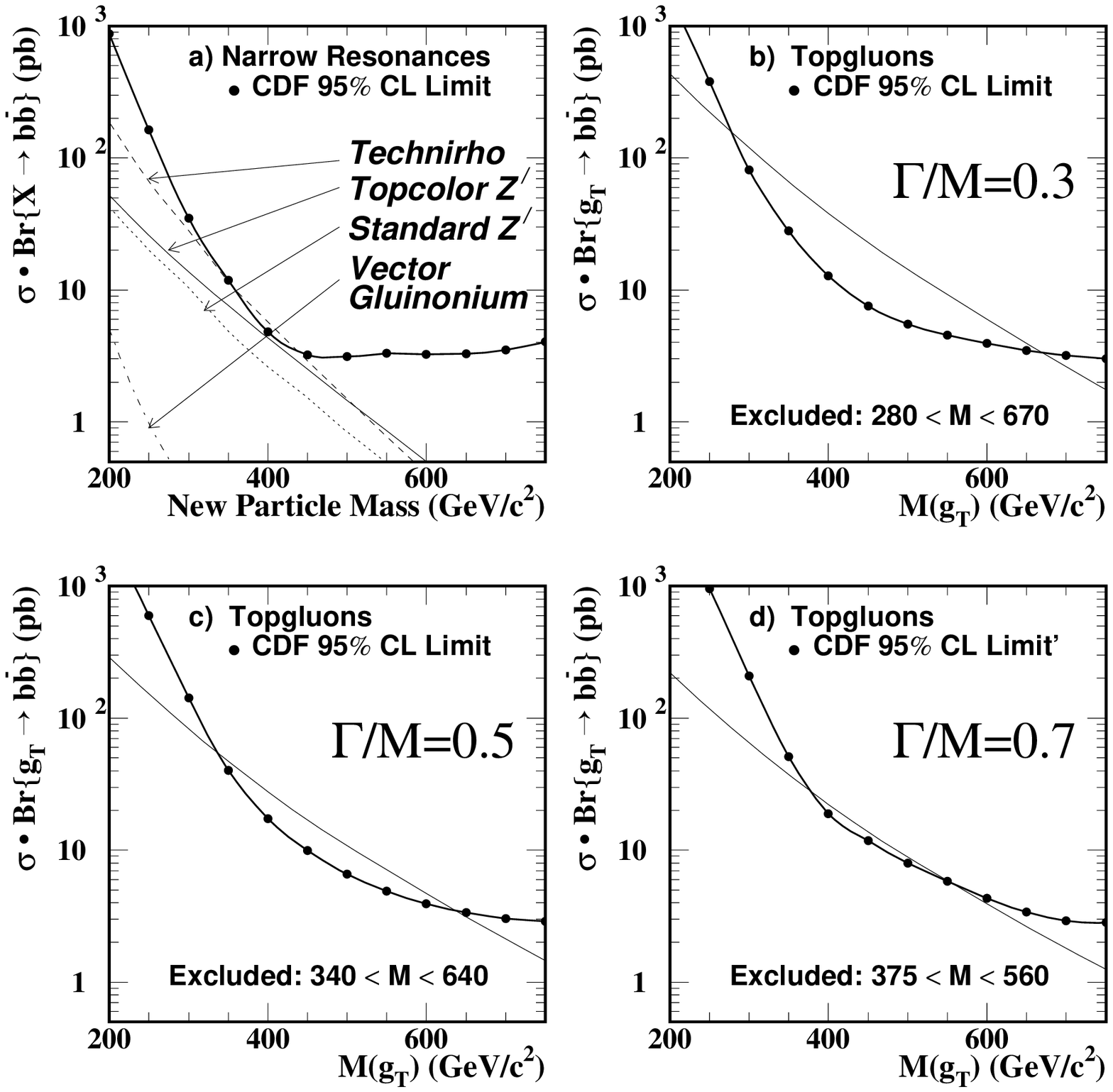,width=6.75in}}
\caption{ 
The 95\% CL upper limit on the cross section times branching ratio 
(points) for a) narrow resonances, and topgluons of width b) $\Gamma=0.3$M, 
c) $\Gamma=0.5$M, and d) $\Gamma=0.7$M is compared to theoretical predictions 
(curves).}
\label{fig_limit}
\end{figure}

In Fig.~\ref{fig_limit} we plot our measured upper limit on the cross section 
times branching ratio as a function of new particle mass in 50 
GeV/c$^2$ steps.  The limit is compared to lowest-order theoretical 
predictions for the cross section times branching ratio for new particles 
decaying to $b\bar{b}$. New--particle decay angular distributions are included 
in the calculations, and we  required $|\eta|<2$ and $|\cos\theta^*|<2/3$ for 
all predictions. 
For narrow resonances we exclude the color 
octet technirho in the mass interval $350 < M < 440$ GeV/c$^2$.  
In addition, our limits for narrow resonances are applicable to any particle 
decaying to $b\bar{b}$ with a width significantly less than our detector resolution of 10\%.
Constructive interference between topgluons and
normal gluons causes the signal cross section to rise at low $b\bar{b}$ mass, 
$m$, and results
in a total cross section integrated over all $m$ that is not well defined.
To avoid this, the total cross section for a topgluon of mass $M$ is defined 
as the cross section in the region $0.5M<m<1.5M$, for both the experimental 
upper limit 
and the theoretical prediction. 
We exclude topgluons of width 
$\Gamma=0.3 M$ in the mass range $280 < M < 670 $ GeV/c$^2$, 
of width $\Gamma = 0.5 M$ in the mass range $340 < M < 640$ GeV/c$^2$, and  
of width $\Gamma = 0.7 M$ in the mass range $375 < M < 560$ GeV/c$^2$. 

In conclusion, the measured $b\bar{b}$ mass spectrum does not contain 
evidence for a mass peak from a new particle resonance. We have 
presented model independent limits on the cross section for a narrow resonance, 
and set specific mass limits on narrow color--octet technirhos and topgluons of 
various widths.

We thank the Fermilab staff and the technical staffs of the participating 
institutions for their vital contributions. This work was
supported by the U.S. Department of Energy and National Science Foundation;
the Italian Istituto Nazionale di Fisica Nucleare; the Ministry of Education, 
Science and Culture of Japan; the Natural Sciences and Engineering Research
Council of Canada; the National Science Council of the Republic of China; 
and the A. P. Sloan Foundation.

\clearpage

\renewcommand{\baselinestretch}{1.4}
\large
\normalsize

\begin{table}
\renewcommand{\baselinestretch}{1.3}
\begin{normalsize}
\begin{center}
\begin{tabular}{|c||c||c|c|c|} \hline\hline
& {\bf Narrow} 
& \multicolumn{3}{|c|}{\bf Topgluons} \\ \hline\hline
Mass & $\Gamma/M <$ 0.1 &$\Gamma/M=0.3$ &
$\Gamma/M=0.5$ & $\Gamma/M=0.7$ \\ \hline
(GeV/c$^2$) & $\sigma$ limit (pb) & $\sigma$ limit (pb) & $\sigma$ limit (pb) 
& $\sigma$ limit (pb) 
\\ \hline\hline
200 & $8.7\times10^2$ & $1.7\times10^3$ & $2.4\times10^3$ & $3.7\times10^3$  \\
250 & $1.6\times10^2$ & $3.8\times10^2$ & $6.0\times10^2$ & $9.6\times10^2$  \\
300 & $3.5\times10^1$ & $8.1\times10^1$ & $1.4\times10^2$ & $2.1\times10^2$  \\
350 & $1.2\times10^1$ & $2.8\times10^1$ & $4.0\times10^1$ & $5.1\times10^1$  \\
400 & $4.8$ & $1.3\times10^1$ & $1.7\times10^1$ & $1.9\times10^1$  \\
450 & $3.2$ & $7.6$ & $9.9$ & $1.2\times10^1$  \\
500 & 3.1   & 5.5  & 6.6  & 8.0 \\ 
550    & 3.3   & 4.5  & 4.9  & 5.8 \\
600    & 3.3   & 4.0  & 3.9  & 4.3 \\
650    & 3.3   & 3.5  & 3.4  & 3.4 \\ 
700    & 3.5   & 3.2  & 3.0  & 2.9 \\
750    & 4.0   & 3.0  & 2.9  & 2.8 \\
\hline\hline
\multicolumn{5}{c}{}\\
\end{tabular}
\caption{The 95\% CL upper limit on the cross section times branching ratio 
for new particles 
decaying to $b\bar{b}$ as a function of new particle mass for narrow 
resonances and for topgluons of three different widths (see text).}
\end{center}
\end{normalsize}
\end{table}

\clearpage

\clearpage

\clearpage

\clearpage


\begin{thebibliography}{99.}

\bibitem{ref_topcolor} C. T. Hill, Phys. Lett. B {\bf 345}, 483 (1995) and 
C. T. Hill and S. J. Parke, Phys. Rev. D {\bf 49}, 4454 (1994).

\bibitem{ref_lane} E. Eichten and K. Lane, New Directions for High-Energy 
Physics, edited by G. Cassel, L. Gennari, and R. Siemann, 1997, page 1006-1009.

\bibitem{ref_zprime} F. Abe {\em et al.}, Phys. Rev. Lett. {\bf 74}, 2900 (1995) and
Phys. Rev. D {\bf 51}, R949 (1995), and references therein.

\bibitem{ref_susy} R. Shanidze, E. Chikovani, V. Kartvelishvili and G. Shaw,
Phys. Rev. D {\bf 53}, 6653 (1994).

\bibitem{ref_CDF} F.\ Abe {\em et al.}, Nucl.\ Instrum.\ and Methods 
{\bf A271}, 387 (1988); F. Abe {\em et al.}, Phys. Rev. D {\bf 50}, 2966 (1994).

\bibitem{ref_jet} F. Abe {\em et al.}, Phys. Rev. D {\bf 45}, 1448 (1992).


\bibitem{ref_top_prl} F. Abe {\em et al.}, Phys. Rev. Lett. {\bf 74}, 2626
(1995).

\bibitem{ref_kara} K. D. Hoffman, Ph.D. thesis, Purdue University, 1998.

\bibitem{ref_pythia} T. Sj\"{o}strand, Comput. Phys. Commun. {\bf 82}, 74 (1994).

\bibitem{ref_CTEQ}  J. Botts {\em et al.}, Phys. Lett. {\bf B304}, 159 (1993).

\end{thebibliography}
\end{document}